\documentclass{llncs}
\usepackage[utf8]{inputenc}
\usepackage[T1]{fontenc}
\usepackage{lmodern}
\usepackage{microtype}
\usepackage{url}
\usepackage[numbers,sectionbib]{natbib}
\usepackage{booktabs}
\usepackage{array}

\usepackage{xcolor}
\usepackage{amsmath}
\usepackage{mathtools}

\let\llncssubparagraph\subparagraph
\let\subparagraph\paragraph
\usepackage{titlesec}
\let\subparagraph\llncssubparagraph
\titlespacing*{\section}{0pt}{12pt}{6pt}
\titlespacing*{\subsection}{0pt}{10pt}{5pt}
\titlespacing*{\subsubsection}{0pt}{5pt}{5pt}

\usepackage[tt=false, type1=true]{libertine} 
\usepackage[varqu]{zi4} 

\usepackage{listings}
\newcommand{\name}{ct-fuzz}

\newcommand{\tup}[1]{{\left\langle{#1}\right\rangle}}

\begin{document}

\pagestyle{plain}
\bibliographystyle{splncs04}
\renewcommand{\bibname}{References}

\title{\name{}: Fuzzing for Timing Leaks}
\author{Shaobo He\inst{1} \and Michael Emmi\inst{2} \and Gabriela Ciocarlie\inst{2}}
\institute{%
University of Utah, Salt Lake City, UT, USA, \email{shaobo@cs.utah.edu}
\and SRI International, New York, NY, USA, \email{\{michael.emmi,gabriela.ciocarlie\}@sri.com}}

\maketitle

\begin{abstract}

  Testing-based methodologies like fuzzing are able to analyze complex software which is not amenable to traditional formal approaches like verification, model checking, and abstract interpretation. Despite enormous success at exposing countless security vulnerabilities in many popular software projects, applications of testing-based approaches have mainly targeted checking traditional safety properties like memory safety. While unquestionably important, this class of properties does not precisely characterize other important security aspects such as information leakage, e.g.,~through side channels.

  In this work we extend testing-based software analysis methodologies to two-safety properties, which enables the precise discovery of information leaks in complex software. In particular, we present the \name{} tool, which lends coverage-guided greybox fuzzers the ability to detect two-safety property violations. Our approach is capable of exposing violations to any two-safety property expressed as equality between two program traces. Empirically, we demonstrate that \name{} swiftly reveals timing leaks in popular cryptographic implementations.

\end{abstract}

\section{Introduction}
\label{sec:intro}

Security is a primary concern for software systems. Programming errors like out-of-bounds memory accesses and inexhaustive input validation are responsible for dangerous and costly incidents. Accordingly, many mechanisms exist for protecting systems against common vulnerabilities like memory-safety errors and input injection. Among the most effective automated approaches is \emph{coverage-based greybox fuzzing}~\citep{DBLP:conf/ccs/BohmePR16, DBLP:conf/ccs/BohmePNR17} popularized by the American Fuzzy Lop ({\sc afl}) fuzzer~\citep{MISC:afl}, which has uncovered copious critical vulnerabilities in the core software libraries underlying a vast swath of software systems. Its efficacy is largely due to broad applicability and direct feedback: {\sc afl} employs a genetic input-generation algorithm using only lightweight instrumentation-based monitoring to determine concrete vulnerability-witnessing inputs. Importantly, it works without user interaction and even source code, and sidesteps the computational and methodological bottlenecks imposed by traditional program analysis techniques.

Fuzzers like afl-fuzz~\citep{MISC:afl} target traditional temporal safety properties~\citep{DBLP:books/daglib/0080029}, i.e.,~properties concerning individual system executions. Important security aspects like secure information flow~\citep{DBLP:journals/jsac/SabelfeldM03} are \emph{two-safety properties}~\citep{DBLP:conf/sas/TerauchiA05}, i.e.,~properties concerning \emph{pairs} of executions, and cannot be precisely characterized as traditional safety properties~\citep{DBLP:conf/sp/McLean94}. Secure information flow is particularly insidious given the potential for \emph{side channels}, through which adversaries may infer privileged information about the data accessed by a program, e.g.,~by correlating execution-time differences with control-flow differences. Side-channels are difficult for programmers to reason about since they are generally hardware-dependent. Furthermore, while compiler optimizations are obliged to respect traditional safety properties, they are not designed to respect two-safety properties, and can thus generate insecure machine code which otherwise would be secure without compiler optimizations.

In this work we extend fuzzing to two-safety properties. In particular, we present \name{},\footnote{\name{}: the fuzzer for constant time. \url{https://github.com/michael-emmi/ct-fuzz}} which enables the precise discovery of timing-based information leaks while retaining the efficacy of afl-fuzz~\citep{MISC:afl}. Via \emph{self-composition}~\citep{DBLP:journals/mscs/BartheDR11}, we reduce testing two-safety properties to testing traditional safety properties by program transformation, effectively enabling the application of any fuzzer. Our implementation tackles three basic challenges. First, the program under test must efficiently simulate execution-pairs of the original; to avoid overhead, each execution’s address space is copy-on-write shared. Second, structured input-pairs must be derived from random fuzzer-provided input; leaks are only witnessed when inputs differ solely by secret content. Third, leakage-inducing actions must be monitored; leaks are witnessed when, e.g.,~control flow or memory-access traces diverge, given inputs differing solely by secret content.

Our implementation focuses on timing leaks related to program control-flow and {\sc cpu} cache; the name \name{} refers to the \emph{constant-time} property, asserting that the control-flow and accessed memory locations of two executions differing solely by secrets are identical. Besides constant-time, we also apply \name{} to finer-grained cache models to validate secure yet non-constant-time programs which ensure identical cache timing despite potentially-divergent memory-access patterns, e.g.,~by \emph{preloading} all accessed memory into the cache. While our implementation focuses on side-channel leakage related to timing, it is extensible to other forms of leakage, e.g.,~power or electro-magnetic radiation, by further extending the instrumentation mechanism to record other aspects of program behavior. In principle, \name{} could be extended to expose violations to any two-safety property expressible as equality between two program traces.

In the remainder we describe several aspects of the \name{} tool. Sections~\ref{sec:theory} and~\ref{sec:design} cover foundations and implementation concerns in testing two-safety properties. Sections~\ref{sec:impl} and~\ref{sec:func} cover \name{}’s software architecture and basic functionality. Sections~\ref{sec:experience} and~\ref{sec:empirical} describe case studies and evaluation, demonstrating the effective application of \name{} to many cryptographic libraries. We survey related and potential future work in Sections~\ref{sec:related} and~\ref{sec:conclusion}.

\section{Theoretical Foundations}
\label{sec:theory}

In this section we characterize \name{}’s foundations, including secure information flow, its reduction to traditional safety properties, and coverage-guided greybox fuzzing.

\subsection{Secure Information Flow}
\label{sec:theory:security}

We consider an abstract notion of secure information flow~\citep{DBLP:journals/jsac/SabelfeldM03} over pairs of executions. We say two executions $e_1$ and $e_2$ are \emph{$f$-equivalent} with respect to some function $f$ when $f(e_1) = f(e_2)$. We model program secrets by a \emph{declassification function} $D$ mapping each execution $e$ to its declassified content $D(e)$, and say $D$-equivalent executions are \emph{comparable}.\footnote{This notion of \emph{declassification} is more general than that appearing in the literature~\citep{DBLP:journals/jsac/SabelfeldM03}.} Intuitively, this equivalence relates executions whose input and output values differ solely by secret content, e.g.,~by the value of a secret key. Similarly, we model attacker capabilities by an \emph{observation function} $O$ mapping each execution $e$ to its observable content $O(e)$, and say two $O$-equivalent executions are \emph{indistinguishable}. Intuitively, this equivalence relates executions which a given observer cannot differentiate, e.g.,~because of negligible timing differences. In practice, we distinguish timing variations by observing divergences in control-flow decisions and accessed memory locations.

\begin{definition}

  A program is \emph{secure} when every pair of comparable executions are indistinguishable, for given declassification and observation functions.

\end{definition}

\subsection{Reducing Security to Safety}
\label{sec:theory:self-comp}

\emph{Self-composition}~\citep{DBLP:journals/mscs/BartheDR11} is program transformation reducing secure information flow to traditional safety properties. Here we say a given program is \emph{safe} when it cannot crash. This simple notion of safety can capture violations to any temporal safety property~\citep{DBLP:books/daglib/0080029} given adequate program instrumentation. For instance, {\sc llvm}’s thread, address, and memory sanitizers signal crashes upon thread- and memory-safety violations, and uses of uninitialized memory, respectively~\citep{DBLP:conf/rv/SerebryanyPIV11, DBLP:conf/usenix/SerebryanyBPV12, DBLP:conf/cgo/StepanovS15}.

\begin{figure}[t]
  \begin{minipage}[b]{0.45\textwidth}
    \centering
    \includegraphics{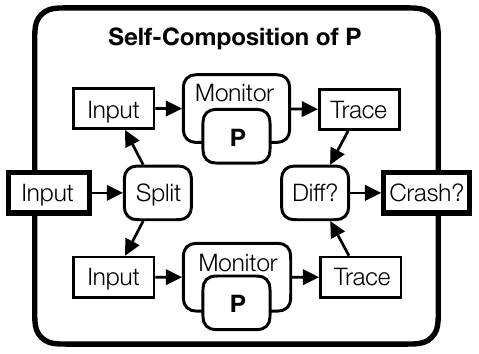}
    \caption{The self-composition simulates two executions of a given program.}
    \label{fig:selfcomp}
  \end{minipage}
  \hfill
  \begin{minipage}[b]{0.45\textwidth}
    \centering
    \includegraphics{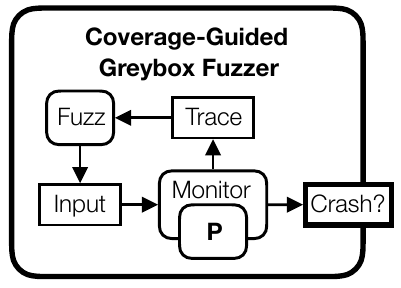}
    \caption{Coverage-guided greybox fuzzers generate inputs and report crashes.}
    \label{fig:fuzzer}
  \end{minipage}
\end{figure}

The \emph{self-composition} is a program (Fig.~\ref{fig:selfcomp}), which executes two copies of a given program in isolation, i.e.,~execution of each copy is independent of the other, which:
\begin{itemize}

  \item halts execution when the copies’ executions are incomparable, and

  \item crashes when the copies’ executions are both comparable and distinguishable.

\end{itemize}
Intuitively, safety of the self-composition implies security of the original program. Conversely, if the original is safe and secure, then the self-composition is also safe.

\begin{theorem}

  A safe program is secure if{f} its self-composition is safe~\citep{DBLP:journals/mscs/BartheDR11}.

\end{theorem}

\subsection{Coverage-Guided Greybox Fuzzing}
\label{sec:theory:fuzzing}

Böhme et al.~\citep{DBLP:conf/ccs/BohmePR16, DBLP:conf/ccs/BohmePNR17} provide a conceptual overview of \emph{coverage-guided greybox fuzzing} as implemented by afl-fuzz~\citep{MISC:afl}. For our purposes, a simplistic view (Fig.~\ref{fig:fuzzer}) suffices: fuzzers explore sequences of program executions to expose safety-property violations (manifested as crashes — see §\ref{sec:theory:self-comp}) by randomly mutating the inputs to previously-explored executions according to per-execution feedback. Intuitively, feedback allows the fuzzer to navigate alternate program paths, and in practice, captures basic-blocks transition counts. From the perspective of this work, the salient aspects of afl-fuzz are its broad applicability and efficiency: it works without user interaction and source code, and avoids prohibitive process-creation overheads by sharing executions’ address spaces in a copy-on-write fashion. These features enable the rapid exploration of program executions for arbitrarily-complex software.

\section{Design and Implementation Concerns}
\label{sec:design}

Applying self-composition (§\ref{sec:theory:self-comp}) to coverage-guided greybox fuzzers (§\ref{sec:theory:fuzzing}) poses three basic challenges: efficiently simulating execution pairs; deriving structured-input pairs from fuzzer-provided inputs (fuzz); and detecting leakage.

\subsection{Simulating Execution Pairs}
\label{sec:design:selfcomp}

The self-composition approach (§\ref{sec:theory:self-comp}) dictates that two identical copies of a program execute in isolation. In the context of testing, achieving isolation includes separating the address spaces of each simulated execution so that the side-effects of one are invisible to the other. On the one hand, the existing approach of duplicating procedures and variables is effective for symbolic analyses like ct-verif~\citep{DBLP:conf/uss/AlmeidaBBDE16}. However, actually executing such duplicated programs on their target platforms can alter the originals’ behavior significantly, e.g.,~due to platform-dependent behavior. On the other hand, executing copies in separate processes undermines the efficacy of fuzzers like afl-fuzz~\citep{MISC:afl}, which largely avoid the overhead of process creation.

\subsection{Deriving Structured Inputs from Fuzz}
\label{sec:design:fuzz}

According to secure information flow (§\ref{sec:theory:security}), leaks are only witnessed by comparable executions, i.e.,~whose declassifications are identical. So on the one hand, testing incomparable executions is wasteful. On the other hand, coverage-guided greybox fuzzers (§\ref{sec:theory:fuzzing}) provide inputs (fuzz) by randomly mutating previously-explored inputs. A self-composition which fed fuzz directly to its simulated executions would generate comparable executions with very low likelihood. For instance, when input variable $x$ is declassified in the program $P(x,y)$, the likelihood of randomly generating the fuzz $\tup{x_1, y_1, x_2, y_2}$ with $x_1 = x_2$ for comparable $P(x_1, y_1)$ and $P(x_2, y_2)$ is extremely low for typical datatypes, e.g., 32- and 64-bit integers. It follows that exposing information leaks requires non-trivial transformations from fuzz to structured input pairs.

\subsection{Detecting Leakage}
\label{sec:design:leaks}

According to secure information flow (§\ref{sec:theory:security}), leaks are witnessed by comparable yet distinguishable executions, i.e.,~whose declassifications are identical, yet observations are distinct. Assuming a reliable mechanism for generating comparable executions (§\ref{sec:design:fuzz}), monitoring leakage amounts to establishing a notion of observation for a given leakage model, instrumenting the source program to record such observations, and signaling a program crash when observations differ.

\section{Software Architecture}
\label{sec:impl}

We implement self-composition by a lightweight {\sc llvm} program transformation~\citep{MISC:LLVM}. We invoke it as part of afl-fuzz’s {\sc llvm}-based instrumentation~\citep{MISC:afl} for convenience, before passing the instrumented program to afl-fuzz. The choice of an {\sc llvm}-based implementation was made for familiarity and convenience, e.g.,~due to {\sc llvm} bitcode being typed; in principle, our approach could be implemented at the assembly level. Following the concerns outlined in Section~\ref{sec:design}, our transformation provides three basic capabilities: efficiently simulating execution pairs, deriving structured inputs-pairs from fuzz, and capturing leakage-relevant observations.

\subsection{Efficient Implementation of Self-Composition}

To implement self-composition, \name{} borrows the same basic trick that makes afl-fuzz efficient: copy-on-write sharing of address spaces via process forks. Figure~\ref{fig:main} sketches our implementation strategy. After initializing the data structures used to capture observations (§\ref{sec:impl:obs}), constructing inputs from fuzz (§\ref{sec:impl:fuzz}), and ensuring the preconditions of the program under test, \name{} forks the running process twice. After each child executes the original program on its copy of input, the parent checks equality of the children’s observation traces. Section~\ref{sec:func:spec} describes the specification of programs, including stating their declassifications and preconditions.

\begin{figure}[t]
	\begin{minipage}[b]{0.475\textwidth}
		\lstset{frame=tb}
\begin{lstlisting}[language=C,morekeywords={pid_t}]
void ct_fuzz_main(void) {
  int status;
  ct_fuzz_initialize();
  ct_fuzz_read_inputs();

  // ensure preconditions
  for (int id = 0; id < 2; ++id)
    ct_fuzz_spec(id);

  // execute each copy
  for (int id = 0; id < 2; ++id) {
    pid_t pid = fork();
    if (pid == -1)
      exit(EXIT_FAILURE);
    else if (pid == 0) {
      ct_fuzz_exec(id);
      exit(EXIT_SUCCESS);
    } else
      waitpid(pid, &status, 0);
  }

  ct_fuzz_check_observations();
}
\end{lstlisting}
		\caption{A fork-based self-composition.}
		\label{fig:main}
	\end{minipage}
	\hfill
	\begin{minipage}[b]{0.475\textwidth}
		\lstset{frame=tb}
\begin{lstlisting}[language=C,morekeywords={bool,num_t}]
// captures observations
// for branch instructions
void ct_fuzz_update_monitor_by_cond(
  bool condition_value,
  char* file_name,
  num_t line_number,
  num_t column_number
);

// captures observations
// for memory accesses
void ct_fuzz_update_monitor_by_addr(
  char* address,
  char* file_name,
  num_t line_number,
  num_t column_number
);
\end{lstlisting}
		\caption{An API for capturing observations.}
		\label{fig:monitor}
	\end{minipage}
\end{figure}

Our fork-based self-composition avoids a potentially-complex address-space management entailed by the existing \emph{duplication} approach (see §\ref{sec:design:selfcomp}). Forking the existing process creates children with identical address spaces, thus minimizing the potential for artificial divergence. Furthermore, the cost of forking is low on modern operating systems due to \emph{copy-on-write} optimization: the virtual-memory pages of child processes point to the same physical pages as their parents; pages themselves are only duplicated when either the parent or child dirties them with subsequent writes. While forking provides ample isolation for basic CPU-driven programs like cryptographic primitives, we do not ensure isolation with stateful IO, e.g.,~interacting with files and sockets; this is a common issue for testing-based approaches.

Our implementation of self-composition also assumes that library functions are deterministic: a sequence of invocations returns identical values in both forked children. Nondeterminism can undermine \name{}’s leakage tests, since divergence between execution pairs may be due to nondeterminism rather than dependence on secrets. This potential is especially apparent in memory allocation: malloc is generally free to return the address of any unallocated chunk of memory. \name{} handles this common case by linking the Jemalloc allocator~\citep{MISC:jemalloc}, which ensures deterministic behavior.

\subsection{Deriving Inputs from Fuzz}
\label{sec:impl:fuzz}

To transform the randomly-mutated inputs (fuzz) into program inputs which are likely to generate comparable executions (see §\ref{sec:design:fuzz}), \name{} generates per-program input processors. These processors depend on the signatures of entry points, as well as \emph{declassification annotations} on program arguments. Besides the program under test, \name{} expects such signatures and annotations to be specified using the API described in Section~\ref{sec:func}. Given this specification, \name{} constructs an input processor, which constructs program-input pairs by reading (from standard input) one fuzzed instance of each declassified argument, and two fuzzed instances of each secret argument. \name{} invokes both program copies with the same fuzz for declassified arguments, and possibly distinct fuzz for secret arguments. While this mechanism does not guarantee that the corresponding executions are ultimately comparable, since further declassification can occur upon execution, e.g.,~declassification of output values, it does avoid incomparability due to declassified inputs.\footnote{While our current implementation does not handle post-input declassification, this could be done by monitoring declassifications, similarly to monitoring observations — see §\ref{sec:impl:obs}.}

\subsection{Recording Observations and Reporting Leakage}
\label{sec:impl:obs}

To capture alternate leakage models \name{} provides an extensible mechanism for recording observations. As our initial implementation targets control-flow and cache-based timing leaks, our program transformation inserts instrumentation before branch and memory-access instructions; more precisely, we insert calls to the monitor API of Figure~\ref{fig:monitor}. The monitor receives branch-condition values and memory addresses, along with the source location of each instruction, and records observations in a shared memory region for access by parent process. When observations diverge, \name{} signals leakage by inducing a crash which will be reported by the fuzzer; our current implementation causes a segmentation fault by dereferencing address zero.

Our initial prototype provides two distinct implementations of the monitor API. The first \emph{constant-time monitor} collects traces of branch-condition values and memory addresses directly; executions are distinguishable if they differ on any of the branch-condition values or memory addresses. The second \emph{cache-model monitor} records Boolean values indicating cache hits or misses in place of memory accesses, allowing for the precise analysis of non-constant-time programs which are nevertheless safe for a given cache architecture, e.g.,~cache-preloading implementations (see §\ref{sec:experience}). Our prototype uses Sung et al.’s cache model~\citep{DBLP:conf/kbse/SungPW18}, though any could be used in its place.

To limit the size of the allocated shared-memory region and time for equality-checking, we leverage a fast hash-function~\citep{MISC:xxHash} to replace the storage of arbitrarily-long observation sequences with one fixed-size hash value. For each observation $o$, the monitor updates its hash value~$m$ to $f(o \cdot m)$ for a given hash function $f$. We state the correctness of this optimization as follows, where a \emph{perfect hash function} $f$ is one in which $f(x) = f(y)$ if{f} $x = y$.

\begin{theorem}

	Given a perfect hash function, two observation sequences are equal if{f} their lengths and corresponding monitors are equal.

\end{theorem}


\section{Functionality and Capabilities}
\label{sec:func}

Our \name{} tool is capable of reporting timing leaks in any code base that can be analyzed with afl-fuzz’s LLVM mode. In this section we demonstrate the use of \name{} on the simple cryptographic function in Figure~\ref{fig:encrypt}.

\begin{figure}[t]
	\lstset{xleftmargin=0.15\textwidth}
	\lstset{xrightmargin=0.15\textwidth}
	\lstset{frame=tb}
\begin{lstlisting}[language=C,morekeywords={aligned}]
// encrypt.c
const char book[10] __attribute__((aligned(64)))
  = { 52, 48, 55, 51, 56, 54, 50, 49, 57, 53 };

void encrypt(char* msg, unsigned len) {
  for (unsigned i = 0; i < len; ++i)
    msg[i] = book[msg[i]-48];
}

// ct-fuzz specification of `encrypt`
#include "ct-fuzz.h"

CT_FUZZ_SPEC(void, encrypt, char* msg, unsigned len) {

  // `msg` is a buffer of length `len` at most `4`
  __ct_fuzz_ptr_len(msg, len, 4);

  // declassification of the `len` argument
  __ct_fuzz_public_in(len);

  // precondition that `msg` contains ASCII decimals
  for (unsigned i = 0; i < len; ++i)
    CT_FUZZ_ASSUME(msg[i]>=48 && msg[i]<=57);
}

CT_FUZZ_SEED(void, encrypt, char*, unsigned) {
  SEED_1D_ARR(char, msg, 4, {'1','2','3','4'})
  PRODUCE(encrypt, msg, 4);
}
\end{lstlisting}
  \caption{A simple (cryptographically-insecure) lookup-table based encryption function.}
  \label{fig:encrypt}
\end{figure}

\subsection{Specifying Argument Preconditions, Declassifications, and Defaults}
\label{sec:func:spec}

Our current implementation of \name{} requires a brief specification of the program under test. Specifically, we require information about program arguments: preconditions, declassifications, and default values. Similarly to ct-verif~\citep{DBLP:conf/uss/AlmeidaBBDE16}, annotations in \name{} are written directly in the source language (i.e.,~C/C++), and processed by our instrumentation at compile time. Specifications are attached to entry points using the {\tt CT\_FUZZ\_SPEC} and {\tt CT\_FUZZ\_SEED} macros. The specification in Figure~\ref{fig:encrypt} declassifies the {\tt len} argument ({\tt msg} is secret), requires that each byte of the {\tt msg} buffer be an {\sc ascii}-encoded decimal digit, and provides a default buffer containing bytes {\tt '1'}–{\tt '4'}.

To facilitate the fuzzing of dynamically-sized buffers like {\tt msg}, the specification also declares that {\tt msg} is a buffer of length {\tt len} at most {\tt 4} bytes; alternatively, constant-length buffers can also be specified. Default values provide the initial values (seeds) from which fuzzers begin their mutation-based exploration. While their specification is not strictly required, since fuzzers can begin with initially-empty seeds, reasonable seeds can boost fuzzers’ performance substantially. We provide macros for specifying default values, e.g.,~{\tt SEED\_1D\_ARR}. Similarly, while preconditions are not strictly required for fuzzing, they allow us to isolate leakage-related crashes from safety-related crashes by assuming that the original program is safe, and thus does not crash, so long as the preconditions are met. The provided {\tt CT\_FUZZ\_ASSUME} macro triggers exit unless its argument evaluates to true. This mechanism leverages the fuzzer’s ability to recognize branches and automatically synthesize inputs which pass precondition checks.

\subsection{Exposure and Diagnosis of Timing Leaks}
\label{sec:func:invoc}

\name{} applies the self-composition described in §\ref{sec:impl} on programs with specifications, generating a binary executable and initial seed for fuzzing. The invocation of \name{}:
\begin{lstlisting}[language=bash]
  $ ct-fuzz --entry-point=encrypt encrypt.c -o encrypt
\end{lstlisting}
generates the {\tt encrypt} binary and seed {\tt 0x31 0x32 0x33 0x34 0x04 0x00 0x00 0x00}, according to the specification of default argument values. After copying the generated seed to {\tt input-dir}, afl-fuzz can be invoked directly:\footnote{Currently we require manual duplication of the seed for the self-composition’s input pair.}
\begin{lstlisting}
  $ afl-fuzz -i input-dir -o output-dir encrypt
\end{lstlisting}
which instantly reports crashes indicating timing leaks, according to the constant-time property. This is the expected result, since the memory locations accessed by the encrypt function depend on the secret contents of {\tt msg}. Specifically, secrets are used as offsets into the {\tt book} buffer. In principle, such secret-dependent memory accesses can lead to cache-based timing variations, and ultimately the leakage of {\tt msg} contents.

To facilitate diagnostics, \name{} includes a mechanism for logging and comparing the observations of the comparable yet distinguishable execution pairs, tracing observations and the source-file locations at which they occur. For example, comparing the traces generated for the leak exposed above:
\begin{lstlisting}
  [dbg] [0] [encrypt.c: 7, 14] [address, 403389]
  ...
  [dbg] [1] [encrypt.c: 7, 14] [address, 403381]
\end{lstlisting}
we spot the divergence due to the dereference of the {\tt book} buffer on Line 7, column 14.

\subsection{Using Alternative Leakage Models}
\label{sec:func:leakage}

For some applications, the constant-time leakage model is too conservative. For example, since the {\tt book} buffer in Fig~\ref{fig:encrypt} fits into a single cache line (assuming standard modern architectures), it would be unlikely that the secret contents of {\tt msg} affect timing, since every access to {\tt book} after the first will hit the cache, independently of {\tt msg}. Alternate leakage models can be selected in \name{} with the {\tt --memory-leakage} flag; currently we support two options: {\tt address} and {\tt cache}. The latter implements Sung et al.’s model~\citep{DBLP:conf/kbse/SungPW18} with fixed values for block size, set associativity, and replacement policy. Extension to parametric and alternative models is straightforward. Using the cache model, we do not discover any timing leaks in the example above.

\section{Experience and Case Studies}
\label{sec:experience}

We have applied \name{} to several popular cryptographic implementations — see §\ref{sec:empirical} for empirical results. To highlight an interesting example, we consider the AES encryption functions of the Botan library invoked in Figure~\ref{fig:code:botan-aes}. For simplicity, we consider potential leakage of a secret 16-byte {\tt key} argument, fixing all other parameters.

\begin{figure}[t]
	\lstset{xleftmargin=0.15\textwidth}
	\lstset{xrightmargin=0.15\textwidth}
	\lstset{frame=tb}
\begin{lstlisting}[language=C++,morekeywords={uint8_t}]
#include <botan/block_cipher.h>
#include <stdint.h>
#include "ct-fuzz.h"

extern "C" {
  void aes_wrapper(uint8_t* key) {
    uint8_t out[64];
    uint8_t data[64] = { 0 };

    std::unique_ptr<Botan::BlockCipher> cipher(
      Botan::BlockCipher::create("AES-128"));

    cipher->set_key(key, 16);
    cipher->encrypt_n(data, out, 4);
  }

  CT_FUZZ_SPEC(void, aes_wrapper, uint8_t* key) {
    __ct_fuzz_ptr_len(key, 16, 16);
  }

  CT_FUZZ_SEED(void, aes_wrapper, uint8_t*) {
    SEED_1D_ARR(uint8_t, key, 16, {42})
    PRODUCE(aes_wrapper, const_cast<uint8_t*>(key));
  }
}
\end{lstlisting}
	\caption{The wrapper program of AES encryption.}
	\label{fig:code:botan-aes}
\end{figure}

\subsection{Analysis of Constant-Time}

Applying \name{} immediately reveals a constant-time violation, witnessed by a pair of inputs differing only by their first byte: {\tt 0xaa} versus {\tt 0x2a}. Comparing their execution traces reveals leakage from the {\tt SE\_word} function shown in Figure~\ref{fig:code:leak}: the single-byte input difference leads to different offsets into the {\tt SE} table computed by {\tt get\_byte(3,x)}.

\begin{figure}[t]
	\lstset{xleftmargin=0.15\textwidth}
	\lstset{xrightmargin=0.15\textwidth}
	\lstset{frame=tb}
\begin{lstlisting}[language=C,morekeywords={inline,uint32_t}]
inline uint32_t SE_word(uint32_t x) {
  return make_uint32(
      SE[get_byte(0, x)],
      SE[get_byte(1, x)],
      SE[get_byte(2, x)],
      SE[get_byte(3, x)]);
}
\end{lstlisting}
	\caption{Secret-dependent array access in Botan AES encryption.}
	\label{fig:code:leak}
\end{figure}

\subsection{Precise Cache Modeling}

Although Botan’s AES implementation is not constant time, it is still considered secure against timing leaks due to its use of \emph{cache preloading} countermeasures. Specifically, every entry in its lookup tables, e.g.,~the {\tt SE} table, is accessed before the secret-dependent lookup-table accesses performed during encryption or decryption to ensure that all subsequent secret-dependent accesses hit the cache. However, previous versions were found to be insecure by SC-Eliminator’s static analyzer~\citep{DBLP:conf/issta/WuGS018} due to missing applications of the countermeasure in the {\tt aes\_key\_schedule} function.\footnote{https://github.com/randombit/botan/commit/09b3d5447d77633d4f9ad0603187ca2a0b017ebd}

Applying \name{} to the insecure version immediately reveals a timing leak even with the cache-model monitor. Interestingly, the inputs for the first-reported leak are identical to those reported in the aforementioned constant-time violation; further analysis demonstrates the same source of leakage in Figure~\ref{fig:code:leak}: one execution’s access of {\tt SE[get\_byte(3,x)]} is a cache miss, while the other’s is a hit. The divergence can be attributed to the fact that the addresses of the four memory accesses in Fig.~\ref{fig:code:leak} are proximate for one execution, while for the other execution, the address of the fourth memory access is distant to the previous three, making them belong to different cache lines. As expected, reapplying \name{} to the secure version reveals no timing leaks.

\section{Empirical Evaluation}
\label{sec:empirical}

We evaluate \name{}’s ability to uncover timing leaks in a range of cryptographic implementations summarized by Table~\ref{tab:implementations}. We analyze multiple entry points from each with the constant-time and cache-model monitors described in Section~\ref{sec:impl}. Besides the benchmarks used to evaluate SC-Eliminator~\citep{DBLP:conf/issta/WuGS018}, we have collected several libraries from their sources. Tables~\ref{tab:experiments:issta} and~\ref{tab:experiments:source} summarize our results.\footnote{We experiment on a 3.5GHz Intel i7 Ubuntu 16.04 desktop machine with 16GB DDR3 memory.} For each entry point, we run afl-fuzz 10 times with 10 second timeouts. In cases where afl-fuzz reports crashes, we re-run afl-fuzz 100 times with the same 10 second timeout, and report the mean and standard deviations until crash for: time (in seconds), number of explored executions, and number of program paths explored. Otherwise, we re-run afl-fuzz 2 times with a 100 second time limit, and report the execution and path counts until timeout.

\begin{table}[t]
	\centering
	\renewcommand{\arraystretch}{1.25}
\begin{tabular}{>{\raggedright\arraybackslash}p{\linewidth}<{}}
  \toprule
  \multicolumn{1}{c}{Implementations taken from original sources} \\
  \midrule
  {\bf BearSSL} v0.5: an implementation of the SSL/TLS protocol (RFC 5246) written in C. \mbox{\url{https://bearssl.org}}
  \\
  {\bf libsodium} {\tt @973cdb5}: a modern, easy-to-use software library for encryption, decryption, signatures, password hashing and more. \mbox{\url{https://download.libsodium.org/doc}}
  \\
  {\bf OpenSSL} v1.1.0h: a robust, commercial-grade, and full-featured toolkit for the Transport Layer Security (TLS) and Secure Sockets Layer (SSL) protocols. \mbox{\url{https://www.openssl.org}}
  \\
  {\bf poly1305-donna} {\tt @e6ad6e0}: a state-of-the-art message-authentication code. \mbox{\url{https://github.com/floodyberry/poly1305-donna}}
  \\
  {\bf poly1305-opt} {\tt @700d5cf}: a portable, performant implementation of Poly1305. \mbox{\url{https://github.com/floodyberry/poly1305-opt}}
  \\
  {\bf s2n} {\tt @745fdd8}: a C99 implementation of the TLS/SSL protocols that is designed to be simple, small, fast, and with security as a priority. \mbox{\url{https://github.com/awslabs/s2n}}
  \\
  \midrule
  \multicolumn{1}{c}{Implementations sourced from Wu et al.~\citep{DBLP:conf/issta/WuGS018}} \\
  \midrule
  {\bf Applied Crypto}: textbook implementations from \emph{Applied Cryptography: protocols, algorithms, and source code in C}~\citep{DBLP:books/daglib/0078909}.
  \\
  {\bf ChronOS}: Cryptographic primitives in ChronOS linux. \mbox{\url{http://www.chronoslinux.org}}
  \\
  {\bf FELICS}: lightweight block ciphers for the Internet of Things. \mbox{\url{https://www.cryptolux.org/index.php/FELICS}}
  \\
  {\bf Libgcrypt}: a general purpose cryptographic library originally based on code from GnuPG. \mbox{\url{https://gnupg.org/software/libgcrypt/index.html}}
  \\
  {\bf SUPERCOP}: a toolkit developed by the VAMPIRE lab for measuring the performance of cryptographic software. \mbox{\url{https://bench.cr.yp.to/supercop.html}}
  \\
  {\bf Botan}: a C++ cryptography library released under the permissive Simplified BSD license. \mbox{\url{https://github.com/randombit/botan/}}
  \\
  \bottomrule
\end{tabular}

	\caption{A list of implementations analyzed with \name{}.}
	\label{tab:implementations}
\end{table}

\begin{table}[t]
\scriptsize
	\begin{minipage}[b]{0.49\textwidth}
		\centering
		\setlength{\tabcolsep}{1pt}
\newcommand{\bm}[1]{{\fontsize{6}{5}\tt #1}}
\begin{tabular}[t]{lcccccc}
  \toprule
  & \multicolumn{3}{c}{Constant Time} & \multicolumn{3}{c}{Cache Model} \\
  \cmidrule(lr){2-4} \cmidrule(lr){5-7}
  Function & Time & Execs & Paths & Time & Execs & Paths \\
  \midrule
  \multicolumn{7}{c}{\bf Botan} \\
  \midrule
  \bm{aes\_key} & 1.1±0.07 & 69±0.87 & 1±0 & 2.1±0.19 & 61±2.3 & 1±0 \\
  \bm{cast128} & 0.35±0.08 & 140±9.3 & 1±0 & 0.38±0.07 & 74±4.2 & 1±0\\
  \bm{des} & 0.36±0.07 & 190±4.8 & 1±0 & 0.36±0.07 & 100±7.4 & 1±0\\
  \bm{kasumi} & 0.36±0.08 & 170±5.1 & 1±0 & 0.37±0.07 & 120±7.3 & 1±0\\
  \bm{seed} & 0.36±0.07 & 160±9.8 & 1±0 & 0.37±0.08 & 99±8.3 & 1±0\\
  \bm{twofish} & 0.6±0.07 & 88±3.1 & 1±0 & 0.59±0.07 & 69±1.4 & 1±0 \\
  \midrule
  \multicolumn{7}{c}{\bf ChronOS} \\
  \midrule
  \bm{aes} & 0.36±0.07 & 160±6.5 & 1±0 & 0.58±0.07 & 140±6.5 & 1±0\\
  \bm{anubis} & 0.35±0.09 & 170±7.4 & 1±0 & 0.58±0.08 & 160±17 & 1±0\\
  \bm{cast5} & 0.77±0.08 & 650±36 & 1±0 & 1.5±0.22 & 610±43 & 1±0\\
  \bm{cast6} & 0.36±0.07 & 220±21 & 1±0 & 0.36±0.09 & 120±14 & 1±0\\
  \bm{des} & 0.36±0.12 & 210±8 & 1±0 & 0.37±0.09 & 130±13 & 1±0\\
  \bm{des3} & 0.36±0.07 & 180±12 & 1±0 & 0.36±0.08 & 110±8.9 & 2.6±0.6\\
  \bm{fcrypt} & 0.36±0.07 & 280±21 & 1±0 & 0.37±0.07 & 200±17 & 1±0\\
  \bm{khazad} & 0.34±0.07 & 270±20 & 1±0 & 0.37±0.1 & 130±9.4 & 1±0\\
  \bottomrule
\end{tabular}

	\end{minipage}
	\hfill
	\begin{minipage}[b]{0.49\textwidth}
		\centering
		\setlength{\tabcolsep}{1pt}
\newcommand{\bm}[1]{{\fontsize{6}{5}\tt #1}}
\begin{tabular}[t]{lcccccc}
  \toprule
  & \multicolumn{3}{c}{Constant Time} & \multicolumn{3}{c}{Cache Model} \\
  \cmidrule(lr){2-4} \cmidrule(lr){5-7}
  Function & Time & Execs & Paths & Time & Execs & Paths \\
  \midrule
  \multicolumn{7}{c}{\bf Applied Crypto} \\
  \midrule
  \bm{loki91} & 0.57±0.08 & 18±0.79 & 1±0 & 0.91±0.10 & 21±1.2 & 1±0 \\
  \bm{3way} & 0.35±0.07 & 190±3.6 & 1±0 & 0.36±0.07 & 150±18 & 1±0\\
  \midrule
  \multicolumn{7}{c}{\bf FELICS} \\
  \midrule
  \bm{LBlock} & 0.36±0.07 & 150±5.2 & 1±0 & -- & 5.4e4±890 & 1±0\\
  \bm{Piccolo} & 0.36±0.08 & 130±9.6 & 1±0 & -- & 4.4e4±1e3 & 1±0\\
  \bm{PRESENT} & 0.37±0.07 & 140±7.6 & 1±0 & -- & 4.3e4±65 & 1±0\\
  \bm{TWINE} & 0.37±0.07 & 88±2.4 & 1±0 & -- & 2.9e4±280 & 1±0 \\
  \midrule
  \multicolumn{7}{c}{\bf Libgcrypt} \\
  \midrule
  \bm{camellia} & 0.35±0.07 & 270±32 & 1±0 & 0.36±0.08 & 190±27 & 1±0\\
  \bm{des} & 0.35±0.07 & 210±19 & 1±0 & 0.35±0.08 & 150±19 & 1±0\\
  \bm{seed} & 0.34±0.07 & 230±40 & 1±0 & 0.37±0.07 & 140±22 & 1±0\\
  \bm{twofish} & 0.37±0.07 & 87±4.3 & 1±0 &0.58±0.15 & 53±12 & 1±0 \\
  \midrule
  \multicolumn{7}{c}{\bf SUPERCOP} \\
  \midrule
  \bm{aes} & 0.35±0.08 & 290±3.2 & 1±0 & 0.36±0.07 & 220±5.8 & 1±0\\
  \bm{cast} & 0.35±0.07 & 230±6.1 & 1±0 & 0.36±0.07 & 100±2.8 & 1±0 \\
  \bottomrule
\end{tabular}

	\end{minipage}
	\caption{Analysis of Wu et al.’s benchmarks~\citep{DBLP:conf/issta/WuGS018} with \name{}. We report averages with standard deviations until leak detection over 100 runs for: time (in seconds), executions, and paths.}
	\label{tab:experiments:issta}
\end{table}

\begin{table}[t]
	\scriptsize
	\centering
	\setlength{\tabcolsep}{2pt}
\newcommand{\bm}[1]{{\tt #1}}
\begin{tabular}{lcccccc}
  \toprule
  & \multicolumn{3}{c}{Constant Time} & \multicolumn{3}{c}{Cache Model} \\
  \cmidrule(lr){2-4} \cmidrule(lr){5-7}
  Function & Time & Execs & Paths & Time & Execs & Paths \\
  \midrule
  \multicolumn{7}{c}{\bf BearSSL} \\
  \midrule
  \bm{aes\_ct(iv,data,{\bf data\_len},key,{\bf key\_len})} & -- & 9.4e4±3.6e3 & 22±0 & -- & 9.3e4±4.4e3 & 30±2.8\\
  \bm{aes\_small({\bf iv},data,{\bf data\_len},key,{\bf key\_len})} & 0.35±0.071 & 1.7e2±0.2 & 1±0 & 1.5±0.13 & 7.5e2±51 & 15±6.8 \\
  \bm{md5(data,{\bf len})} & -- & 1.3e5±3.9e3 & 2±0 & -- & 1.3e5±2.2e3 & 2±0 \\
  \midrule
  \multicolumn{7}{c}{\bf libsodium} \\
  \midrule
  \bm{\shortstack{crypto\_aead\_chacha20poly1305\_encrypt(\\
    c,clen,m,{\bf mlen},{\bf ad},{\bf adlen},{\bf npub},k)}} & -- & 4.2e4±1.5e3 & 31±0 & -- & 3.5e4±3.6e2 & 46±2.1 \\
  \bm{crypto\_shorthash(out,in,{\bf inlen},k)} &  -- & 1.3e5±5.5e3 & 11±0 & -- & 1.2e5±8.8e2 & 13±1.4 \\
  \bm{sodium\_increment(n,{\bf nlen})} & -- & 1.4e5±7.2e3 & 12±0 & -- & 1.4e5±8.8e2 & 16±0.7\\
  \bm{sodium\_is\_zero(n,{\bf nlen})} & -- & 1.4e5±3.4e3 & 8±0 & -- & 1.4e5±4.4e3 & 12±2.1\\
  \midrule
  \multicolumn{7}{c}{\bf OpenSSL} \\
  \midrule
  \bm{EVP\_aes\_128\_cbc(key,data,{\bf iv})} & 0.36±0.08 & 2.5e2±7.7 & 1±0 & 0.34±0.07 & 2e2±1.6 & 1±0\\
  \bm{ssl3\_cbc\_copy\_mac({\bf orig\_len\_raw},length)} & -- & 2.8e5±5.3e3 & 7.5±0.7 & -- & 2.7e5±1.1e3 & 8.5±0.7\\
  \bm{ssl3\_cbc\_copy\_mac\_modulo({\bf orig\_len\_raw},length)} & -- & 2.6e5±8.9e3 & 8.5±0.7 & -- & 2.7e5±3.7e3 & 10±0\\
  \midrule
  \multicolumn{7}{c}{\bf poly1305-donna} \\
  \midrule
  \bm{poly1305\_auth(mac,m,{\bf bytes},key)} & -- & 1.3e5±1.8e3 & 15±0 & -- & 1.4e5±2.8e3 & 24±0.7 \\
  \midrule
  \multicolumn{7}{c}{\bf poly1305-opt} \\
  \midrule
  \bm{poly1305\_auth(mac,in,{\bf inlen},key)} & -- & 2.6e4±2.2e2 & 1±0 & -- & 2.4e4±4.4e2 & 2±1.4 \\
  \midrule
  \multicolumn{7}{c}{\bf s2n} \\
  \midrule
  \bm{s2n\_hmac\_digest(sekrit,{\bf sekritlen},msg,{\bf msglen})} & -- & 9.3e4±8.7e2 & 4±0 & -- & 8.2e4±2.5e3 & 4.5±0.7 \\
  \bottomrule
\end{tabular}

	\caption{Analysis of open-source crypto implementations with \name{}. Declassified inputs are listed in boldface. We report averages with standard deviations until leak detection over 100 runs for: time (in seconds), executions, and paths.}
	\label{tab:experiments:source}
\end{table}

\subsection{Analysis with the Constant-Time Monitor}

\name{} swiftly reports constant-time violations due to secret-dependent table lookups, e.g.,~all of Wu et al.’s benchmarks. OpenSSL's C implementation of AES encryption leverages substitution boxes. For supposedly constant-time implementations such as BearSSL's constant-time AES encryption (\texttt{aes\_ct}) and libsodium's constant-time utility functions (\texttt{sodium\_increment}, \texttt{sodium\_is\_zero}), \name{} reports no violations. In our experience with these benchmarks, afl-fuzz does not report any violations, even after several hours, that are not reported within a few seconds. On average, \name{} uncovers violations within half of a second, exploring only hundreds of executions. Execution counts are approximate, since afl-fuzz’s {\tt AFL\_BENCH\_UNTIL\_CRASH} mode only guarantees termination \emph{soon} after the first crash. Furthermore, path counts report the number of unique non-crashing control-flow paths discovered by afl-fuzz; paths are often unique since cryptographic implementations tend to use fairly straight-line code.

\subsection{Analysis with the Cache-Model Monitor}

We further run \name{} with our cache-model monitor (see §~\ref{sec:func:leakage}). The cache model is identical to that used in the evaluation of SC-Eliminator~\citep{DBLP:conf/issta/WuGS018}: a fully associative LRU cache of 512 64-byte lines. We report no timing leaks for Wu et al.’s FELICS benchmarks~\citep{DBLP:conf/issta/WuGS018} according to the cache-model monitor, which is expected since their lookup tables are small enough to fit into one cache line. This is consistent with SC-Eliminator's static analysis, although our notion of leakage, framed as a two-safety property, is a more-precise characterization which avoids the potential for false positives — see §\ref{sec:related}. Furthermore, we report no timing leaks for constant-time functions like {\tt aes\_ct}, which is expected since constant-time is stricter than the absence of leakage according to a precise cache model. The differences in the number of executions explored until crash compared with the constant-time monitor is due to the approximation of execution counts: since the constant-time monitor is more efficient than the cache-model monitor, afl-fuzz squeezes in many more executions between the first crash and termination.

\section{Related Work}
\label{sec:related}

The literature around secure information flow is vast, and Sabelfeld and Myers offer a compelling survey~\citep{DBLP:journals/jsac/SabelfeldM03}. Here we highlight recent works to detecting timing side-channel leaks with program analysis. We distinguish between \emph{symbolic} approaches, e.g.,~involving verification, model checking, and abstract interpretation, and \emph{concrete} approaches, e.g.,~testing and coverage-guided greybox fuzzing, which observe executions on programs’ target platforms. Along another axis, we differentiate works according to how they identify leakage, whether it be \emph{tracking the flow of data} through program values, \emph{statistical measures} on the distributions of program executions, or \emph{as two-safety properties} on pairs of program executions.

\subsection{Symbolic Analyses}

While our work accurately \emph{detects} leakage, most related works soundly establish the \emph{absence} of leaks. In contrast, our testing-based approach sacrifices soundness\footnote{While coverage-guided fuzzing is subject to false negatives, e.g.,~potential leakage not reported, we have found that all potential leakages are swiftly reported in typical cryptographic implementations since control- and data-flow is straightforward.} for precision and efficiency, without introducing the potential for false positives or computational bottlenecks due to abstract symbolic reasoning. We identify several works based on symbolic analysis of data-flow or two-safety properties.

\subsubsection{Flow Tracking}

Many recent works propose static analyses to prove secure information flow. tis-ct~\citep{MISC:TIS-CT} extends Frama-C~\citep{DBLP:conf/icfp/CuoqSBBCCMPP09} with static analysis to compute program dependencies and infer potential leaks. VirtualCert~\citep{DBLP:conf/ccs/BartheBCLP14} integrated a type system capable of tracking aliasing and information flow into the CompCert certified compiler~\citep{DBLP:journals/jar/Leroy09}. FlowTracker~\citep{DBLP:conf/cc/RodriguesPA16} developed efficient static information-flow analyses based on the single static assignment ({\sc ssa}) representation rather than program dependence graphs ({\sc pdg}). Blazy et al.~\citep{DBLP:conf/esorics/BlazyPT17} developed context-sensitive and arithmetic-aware alias analyses to verify constant time of C programs using the Verasco static analyzer~\citep{DBLP:conf/popl/JourdanLBLP15}. CacheAudit~\citep{DBLP:journals/tissec/DoychevKMR15,DBLP:conf/pldi/DoychevK17} developed effective cache-aware abstract domains to prove the absence of cache-based side channels using abstract interpretation. SC-Eliminator~\citep{DBLP:conf/issta/WuGS018} performs static analysis to identify or prove the absence of potential cache-timing leaks, and synthesizes patches by replacing conditional statements and lookup-table accesses.

Other works employ satisfiability-based techniques. CacheD~\citep{DBLP:conf/uss/WangWLZW17} uses symbolic execution to identify potential cache-access variations as expressions over secrets. SCInfer~\citep{DBLP:conf/cav/ZhangGSW18} combines type-inference and satisfiability modulo theories ({\sc smt}) to verify random masking of program secrets. {\sc canal}~\citep{DBLP:conf/kbse/SungPW18} performs analysis-friendly program instrumentation enabling symbolic reasoning about cache-related properties, e.g.,~which accesses are cache hits. {\sc cachefix}~\citep{DBLP:journals/tcad/ChattopadhyayR18} uses {\sc cbmc}~\citep{MISC:CBMC} to prove the absence of leaks, or synthesizes patches for discovered leaks.

\subsubsection{Two-Safety}

Barthe et al.~\citep{DBLP:journals/mscs/BartheDR11} originally proposed the \emph{self-composition} program transformation to reduce the verification of \emph{two-safety} properties like secure information flow to traditional safety verification. Terauchi and Aiken~\citep{DBLP:conf/sas/TerauchiA05} refined their approach with a type-directed transformation to apply self-composition selectively, with a simpler \emph{cross-product} transformation~\citep{DBLP:conf/fm/ZaksP08} around low-security program fragments. Milushev et al.~\citep{DBLP:conf/forte/MilushevBC12} adapt this approach for use with the {\sc klee} symbolic execution engine~\citep{DBLP:conf/osdi/CadarDE08}, while ENCoVer~\citep{DBLP:conf/csfw/BalliuDG12} leverages Java PathFinder~\citep{DBLP:conf/issta/VisserPK04}. Phan~\citep{DBLP:conf/iccsw/Phan13} reformulates self-composition as \emph{path equivalence} for direct symbolic execution of the original program.

Almeida et al.~\citep{DBLP:journals/scp/AlmeidaBPV13} developed a self-composition based methodology for proving \emph{constant-time} of C programs using Frama-C~\citep{DBLP:conf/icfp/CuoqSBBCCMPP09}. The recently developed ct-verif~\citep{DBLP:conf/uss/AlmeidaBBDE16} and SideTrail~\citep{MISC:conf/vstte/AthanasiouCEMST18} tools apply selective self-composition to verify constant-time and time-balancing, respectively, in C-language cryptographic primitives using the {\sc smack} verifier~\citep{DBLP:conf/cav/RakamaricE14}. Yang et al.~\citep{DBLP:conf/cav/YangVSGM18} propose \emph{lazy self-composition} to apply precise reasoning only as a fallback to coarser techniques like taint analysis. Blazer~\citep{DBLP:conf/pldi/AntonopoulosGHK17} proposes \emph{quotient partitioning} as an alternative, compositional reduction to traditional safety verification.

\subsection{Concrete Analyses}

More closely aligned with our approach are those based on observing executions on programs’ target platforms, thus avoiding the potential for false positives and computational bottlenecks due to abstract symbolic reasoning. While we consider secure information flow as a two-safety property, other works are based either on tracking data-flow or statistical analysis.

\subsubsection{Flow Tracking}

The key work in this space is ctgrind~\citep{MISC:ctgrind}, which extends Valgrind~\citep{MISC:Valgrind} with a taint-tracking mechanism to track whether secrets flow into accessed memory locations. While this avoids the bottlenecks of symbolic reasoning, this approach is susceptible to false positives, since secret-tainted observations are falsely considered leaks when they depend also on public or declassified data values. Considering secure information flow as a two-safety property avoids this source of false positives.

\subsubsection{Statistics}

A few approaches detect timing leaks using statistical methods. dudect~\citep{DBLP:conf/date/ReparazBV17} records the execution time of many (typically millions) of executions with inputs partitioned into random and fixed values. While measuring execution time directly avoids the imprecision of indirect measurements like memory-access traces, this can also be considered a source of false negatives, since timing observed in a given test environment can vary significantly from other platforms and system loads.

Other statistical approaches measure traces of memory and control-flow operations. {\sc data}~\citep{DBLP:conf/uss/WeiserZSMMS18} detects differences among traces of executions with random inputs, before honing in on statistical tests with fixed- versus random-secret inputs; the approach relies on expert interaction for classification of potential leaks, and is susceptible to false positives due to program nondeterminism, e.g.,~of memory allocators. MicroWalk~\citep{DBLP:journals/corr/abs-1808-05575} proposes alternative statistical tests based on \emph{mutual information} analysis.

\subsubsection{Two-Safety}

To the best of our knowledge, {\sc mutaflow}~\citep{DBLP:conf/se/MathisASBZ18} is the only other concrete-analysis approach considering secure information flow as a two-safety property. Like our work, {\sc mutaflow} compares observations along two executions in which the second receives the first’s inputs randomly mutated. However, since only program values are observed, {\sc mutaflow} cannot detect leakage due to side-channels like timing.

\section{Conclusion and Future Work}
\label{sec:conclusion}

This work demonstrates that testing-based software analysis methodologies can be effectively extended from safety properties to two-safety properties. While our initial application has targeted timing leaks in cryptographic implementations, there are several promising directions for future investigation. Since cryptographic code tends to follow rather simple straight-line control flow, coverage-guided greybox fuzzers offer relatively little over more simplistic random testing; exposing leaks in more complex input-processing applications such as {\sc jpeg} encoders and decoders~\citep{DBLP:conf/sp/XuCP15} could better exploit two-safety-property fuzzers. Furthermore, \name{} could be applied to other types of leakage, and even other classes of two-safety properties, by extending the instrumentation mechanism; in principle, our approach is capable of exposing violations to any two-safety property expressed as equality between program traces.

\section*{Acknowledgements}

This work was funded in part by the US Department of Homeland Security (DHS) Science and Technology (S\&T) Directorate under contract no.~HSHQDC-16-C-00034. The views and conclusions contained herein are the authors’ and should not be interpreted as necessarily representing the official policies or endorsements, either expressed or implied, of DHS or the US government.

\bibliography{dblp,misc}

\end{document}